\documentclass[copyright,creativecommons]{eptcs}

\usepackage{graphicx} 

\usepackage{times}

\usepackage{amsmath}

\usepackage{hyperref}

\usepackage{cleveref}

\usepackage{comment}

\usepackage{graphicx}
\usepackage[font=small,labelfont=bf,margin=\parindent,tableposition=top]{caption}
\usepackage{subcaption}
\usepackage{appendix}
\usepackage{stmaryrd,paralist} 

\usepackage{textcomp}
\usepackage{latexsym}
\usepackage{amsthm}

\DeclareMathAlphabet{\mathbb}{U}{msb}{m}{n}
\usepackage{MnSymbol}
\usepackage{relsize}
\usepackage{mathtools}

\usepackage{tabularx}

\usepackage{wrapfig}

\usepackage{listings}

\usepackage{tikz}
\usetikzlibrary{datavisualization}
\usetikzlibrary{datavisualization.formats.functions}
\usetikzlibrary{calc,trees,positioning,arrows,chains,shapes.geometric,
    decorations.pathreplacing,decorations.pathmorphing,shapes,
    matrix,shapes.symbols}  
\usepackage{pgf}
\usepackage{pgfplots}    
\usepackage{adjustbox}
\usepackage{booktabs}
\usepackage{multirow}
\usepackage{diagbox}
\usepackage{makecell}

\pgfplotsset{compat=1.17}

\newcommand*{\mymodels}{\: \mathlarger{\models} \:}
\newcommand{\currenttime}{\tau}

\newtheorem{remark}{Remark}
\newtheorem{property}{Property}
\newtheorem{definition}{Definition}

\usepackage{todonotes}

\lstdefinelanguage{stateflow}{
  morekeywords={if,then,else,let,in,match}
}

\definecolor{cdarkgreen}{rgb}{0.0,0.4,0.0}
\definecolor{customblue}{rgb}{0.0,0.0,0.7}
\definecolor{cbluegreen}{rgb}{0.0,0.4,0.7}
\definecolor{cpurple}{rgb}{0.5,0.0,0.7}
\definecolor{corange}{rgb}{0.8,0.6,0.2}
\definecolor{cgreen}{rgb}{0,0.6,0}
\colorlet{commentcolour}{green!50!black}
\colorlet{keywordcolour}{magenta!90!black}
\definecolor{MyDarkGreen}{rgb}{0.0,0.4,0.0}
\definecolor{MyBlue}{rgb}{0.0,0.0,0.7}
\definecolor{MyPurple}{rgb}{0.7,0.0,0.7}
\usepackage{tikz}
\usepackage[framemethod=TikZ]{mdframed}
\lstdefinestyle{acslstyle}{language=C++,                
  alsoletter={<==>,==>,\&\&,||,\\,->},
  keywords = [2]{assumes,requires,ensures,behavior,logic,axiomatic,type,predicate,axiom,<==>,==>,\&\&,||,assert,loop, invariant, variant, ghost, for},
  keywords = [3]{\\valid,\\separated,\\forall,\\result,\\exact,\\model,\\total_error,\\error,\\abs,\\cos,\\sin,\\set_model,\\sqrt},
  keywords = [4]{PROOF_TACTIC},
  keywordstyle=[1]\color{blue},
  keywordstyle=[2]\color{Purple},
  keywordstyle=[3]\color{BlueGreen},
  keywordstyle=[4]\color{BurntOrange},
  morecomment=[l][\color{OliveGreen}]{//},
  morecomment=[s][\color{OliveGreen}]{/*}{*/},
  moredelim=**[s][\color{OliveGreen}]{/*@}{*/},
  moredelim=**[l][\color{OliveGreen}]{//@},
  rulecolor=\color{black},
  alsoother = ,}

\lstdefinestyle{pvsstyle}{morecomment=[l]{\%},
    commentstyle=\color{Maroon},
    keywords = [2]{lambda,type,forall,conversion,var,if, then, else,
      endif, importing, begin, theory, end,
      auto_rewrite,lemma,theorem,macro,let,implies,and,in,table,endtable,axiom,iff,true,false},
    sensitive=false,
    keywordstyle = [2]\color{Purple},
    literate=*
    {@}{{{\color{blue}+}}}1
    {+}{{{\color{blue}+}}}1
    {*}{{{\color{blue}*}}}1
    {=}{{{\color{blue}=}}}1
    {:=}{{{\color{blue}:=}}}2
    {/}{{{\color{blue}/}}}1
    {-}{{{\color{blue}-}}}1
    {<}{{{\color{blue}<}}}1
    {>}{{{\color{blue}>}}}1
    {<=}{{{\color{blue}<=}}}2
    {>=}{{{\color{blue}>=}}}2
    {->}{{{\color{black}->}}}2}

\lstdefinestyle{lustrestyle}
    {
  morekeywords={[1]guarantee,assume,contract,PROPERTY,by,induction},
  morekeywords={[2]int,bool,float,double,real,clock},
  morekeywords={[3]true,false},
  morekeywords={[4]node,contract,returns,var,let,tel,automaton,state,resume,restart,unless,until,type,enum},
  morekeywords={[5]fby,when,whenot,merge,every,not,and,or,pre,if,then,else},
  sensitive=true, 
  moredelim=**[s][\color{blue}]{(*}{*)},
  moredelim=**[l][\color{darkgray}]{--@},
  moredelim=**[l][\color{darkgray}]{--\%},
  moredelim=**[l][\color{blue}]{--},
    }[keywords,comments]

\lstdefinestyle{amsl}{language=Matlab,
breakatwhitespace=false,    
breaklines=true, 
postbreak=\raisebox{0ex}[0ex][0ex]{\ensuremath{\color{red}\hookrightarrow\space}}, 
rulecolor=\color{black!40},
emphstyle=\color{blue},
keywordstyle=\color{keywordcolour}\bfseries,
commentstyle=\color{commentcolour}\slshape,
morekeywords = {length,lsqr,norm,transpose,trace}, 
keywords = [2]{assumes,requires,ensures,behavior,predicate,logic,ghost,type,\&\&,||,loop, invariant}, 
keywords = [3]{PROOF_TACTIC},
keywords = [4]{on, off}, 
keywords = [5]{vecs,mats,smat,krons,is_symmetric,is_matrix,is_vector,type}, 
keywordstyle=[1]\color{MyBlue}, 
keywordstyle=[2]\color{cpurple},
keywordstyle=[3]\color{corange},
keywordstyle=[4]\color{MyPurple}, 
keywordstyle=[5]\color{cbluegreen}, 
morecomment=[l][\color{cgreen}]{//},
morecomment=[s][\color{cgreen}]{/*}{*/},
moredelim=**[s][\color{cgreen}]{/*@}{*/},
moredelim=**[l][\color{cgreen}]{//@},
}
\lstdefinestyle{matl}{language=Matlab,
breakatwhitespace=false,    
breaklines=true, 
rulecolor=\color{black!40},
emphstyle=\color{blue},
keywordstyle=\color{keywordcolour}\bfseries,
commentstyle=\color{commentcolour}\slshape,
morekeywords = {length,lsqr,norm,transpose,trace}, 
keywords = [2]{assumes,requires,ensures,behavior,predicate,logic,ghost,type,\&\&,||,loop, invariant}, 
keywords = [3]{PROOF_TACTIC},
keywords = [4]{on, off}, 
keywords = [5]{vecs,mats,smat,krons,is_symmetric,is_matrix,is_vector,type}, 
keywordstyle=[1]\color{MyBlue}, 
keywordstyle=[2]\color{cpurple},
keywordstyle=[3]\color{corange},
keywordstyle=[4]\color{MyPurple}, 
keywordstyle=[5]\color{cbluegreen}, 
morecomment=[l][\color{cgreen}]{//},
morecomment=[s][\color{cgreen}]{/*}{*/},
moredelim=**[s][\color{cgreen}]{/*@}{*/},
moredelim=**[l][\color{cgreen}]{//@},
}

\makeatletter
\def\mdf@@codeheading{Code Listings}
\define@key{mdf}{title}{%
   \def\mdf@@codeheading{#1}
}
\mdfdefinestyle{lstlisting}{%
 innertopmargin=5pt,
 innerbottommargin=0pt,
 middlelinewidth=1pt,
 outerlinewidth=9pt,
 outerlinecolor=white,
 innerleftmargin=2pt,
 innerrightmargin=2pt,
 leftmargin=-9pt,rightmargin=-9pt,
 skipabove=\topskip,
 skipbelow=\topskip,
 roundcorner=5pt,
 singleextra={\node[draw, fill=white,anchor=east, rounded corners=5pt,
     xshift=410pt+1pt] at (O|-P) {\csname mdf@@codeheading\endcsname};},
 firstextra={\node[draw, fill=white,anchor=east, rounded corners=5pt,
     xshift=210pt+1pt] at (O|-P) {\csname mdf@@codeheading\endcsname};}
}
\makeatother

\lstnewenvironment{lustre}{%
  \lstset{
    style=lustrestyle,
    mathescape,
    basicstyle=\small\ttfamily,
    numbers=left, 
    stepnumber=1, 
  }
}{
}

\lstnewenvironment{acsl}{%
  \lstset{style=acslstyle,mathescape}%
  \mdframed[style=lstlisting,title={ACSL}]%
}{\endmdframed}

\lstnewenvironment{python}{%
  \lstset{style=acslstyle,mathescape,language=python}%
  \mdframed[style=lstlisting,title={Python}]%
}{\endmdframed}

\lstnewenvironment{cacsl}{%
  \lstset{style=acslstyle,mathescape}%
  \mdframed[style=lstlisting,title={C+ACSL}]%
}{\endmdframed}

\lstnewenvironment{ccode}{%
  \lstset{style=acslstyle,mathescape}%
 \mdframed[style=lstlisting,title={C},nobreak=true]%
 \vspace{-.8em}
}{\endmdframed}

\lstnewenvironment{pvs}{%
  \lstset{style=pvsstyle,mathescape}%
  \mdframed[style=lstlisting,title={PVS}]%
}{\endmdframed}

\lstnewenvironment{matlab}{%
  \lstset{style=amsl,mathescape}%
 \mdframed[style=lstlisting,title={Matlab},nobreak=true]%
 \vspace{-.8em}
}{\endmdframed}

\newcommand{\lustrein}{\lstinline[style=lustrestyle,mathescape,basicstyle=\ttfamily]}

\lstnewenvironment{ccodenum}{%
  \lstset{style=acslstyle,mathescape,numbers=left,xleftmargin=2em}%
  \mdframed[style=lstlisting,title={C},nobreak=true]%
 \vspace{-.8em}
}{\endmdframed}

\definecolor{module}{cmyk}{0.4,1,0.3,0.5}
\definecolor{constructor}{cmyk}{0,0.9,0,0.3}
\definecolor{function}{rgb}{0.1,0.65,0.1}
\definecolor{type}{rgb}{0.8,0,0}
\definecolor{variable}{rgb}{0,0,1}

\lstnewenvironment{ocaml}{%
  \lstset{language=[Objective]Caml,
  alsoletter={\textunderscore},
  basicstyle=\ttfamily,
  keywordstyle=\bfseries,
  moredelim=[is][\color{variable}]{//}{//},
  moredelim=[is][\color{type}]{!!}{!!},
  moredelim=[is][\color{function}]{--}{--},
  emphstyle=[1]{\color{variable}},
  emphstyle=[2]{\color{module}},
  emphstyle=[3]{\color{constructor}},
  emphstyle=[4]{\color{function}},
  emphstyle=[5]{\color{type}},
  identifierstyle=\color{variable},
  commentstyle=\itshape\color{gray},
 basewidth=0.45em,
  emph=[3]{None,Some},
  emph=[5]{float,int,bool,ref,unit,option},
}%
  \mdframed[style=lstlisting,title={OCaml}]%
}{\endmdframed}

\colorlet{punct}{red!60!black}
\definecolor{background}{HTML}{EEEEEE}
\definecolor{delim}{RGB}{20,105,176}
\colorlet{numb}{magenta!60!black}

\lstdefinelanguage{json}{
    basicstyle=\normalfont\ttfamily,
    numberstyle=\scriptsize,
    numbersep=8pt,
    showstringspaces=false,
    breaklines=true,
    frame=lines,
    backgroundcolor=\color{background},
    literate=
     *{0}{{{\color{numb}0}}}{1}
      {1}{{{\color{numb}1}}}{1}
      {2}{{{\color{numb}2}}}{1}
      {3}{{{\color{numb}3}}}{1}
      {4}{{{\color{numb}4}}}{1}
      {5}{{{\color{numb}5}}}{1}
      {6}{{{\color{numb}6}}}{1}
      {7}{{{\color{numb}7}}}{1}
      {8}{{{\color{numb}8}}}{1}
      {9}{{{\color{numb}9}}}{1}
      {:}{{{\color{punct}{:}}}}{1}
      {,}{{{\color{punct}{,}}}}{1}
      {\{}{{{\color{delim}{\{}}}}{1}
      {\}}{{{\color{delim}{\}}}}}{1}
      {[}{{{\color{delim}{[}}}}{1}
      {]}{{{\color{delim}{]}}}}{1},
}
\lstdefinestyle{jsonstyle}{language=json,                
}

\lstnewenvironment{json}{%
  \lstset{style=jsonstyle,mathescape,basicstyle=\linespread{1.0}\small\ttfamily}
}{
}

\newcommand{\therepository}{\url{https://garoche.net/publication/2023_fmas_submission/}}

\title{Towards Proved Formal Specification and Verification of STL Operators as Synchronous Observers}
\author{Céline Bellanger\institute{ENAC, Université de Toulouse}
\and  Pierre-Loïc Garoche\institute{ENAC, Université de Toulouse}
\and Matthieu Martel\institute{Université de Perpignan Via Domitia}
\and Célia Picard\institute{ENAC, Université de Toulouse}}
\date{}
\def\titlerunning{STL Operators as Synchronous Observers}
\def\authorrunning{C. Bellanger, P.-L. Garoche, M. Martel \& C. Picard}

\begin{document}

\maketitle

\begin{abstract}

    Signal Temporal Logic (STL) is a convenient formalism to express bounded horizon properties of autonomous critical systems. STL extends LTL to real-valued signals and associates a non-singleton bound interval to each temporal operators. In this work we provide a rigorous 
    encoding of non-nested discrete-time STL formulas into Lustre synchronous observers. 
    
    Our encoding provides a three-valued online semantics for the observers and therefore enables both the verification of the property and the search of counter-examples. A key contribution of this work is an instrumented proof of the validity of the implementation. Each node is proved correct with respect to the original STL semantics. All the experiments are automated with the Kind2 model-checker and the Z3 SMT solver.

\end{abstract}

\section{Introduction} \label{sec:intro}

In the context of autonomous critical systems, an undesirable behaviour can lead to significant material or human damage. Thus, the specification of properties and their formal verification play a paramount role in ensuring the safety, reliability and compliance of such systems. 

Dynamical systems continuously respond to environmental changes. Signal Temporal Logic (STL) has emerged as a powerful formalism for expressing temporal properties within these systems~\cite{maler_monitoring_2004}. The main particularity of STL language is the association of each temporal operator with a finite, non-singleton time interval, during which the operator is studied. Consider the temporal $\Diamond$ \textit{(Eventually)} operator, which evaluates whether a property $\varphi$ is satisfied or not at least once. A correct formalism for $\Diamond$ in STL is $\Diamond_{[a,b]} \varphi$, where $a$ and $b$ are times such that $a < b$. 
Most of the time, STL properties are assess offline: we execute the system from start to finish, and we observe after the end of the execution if the system behaviour and its outputs are compliant to specified requirements. 

However, the complexity of certain autonomous dynamical systems may require runtime verification. This involves continuous assessment of the system's compliance to its specification throughout execution. Synchronous observers can be employed for this purpose. These specialized observers react when a property is satisfied or violated, providing instantaneous information about the system's state. This approach offers advantages such as consistent real-time information transmission and the ability to halt executions immediately upon property satisfaction or violation, without waiting for completion. Notably, this enables quicker reactions to external events, crucial for critical systems. For instance, let us admit that we wish to satisfy a property $\varphi$ at least once during a time interval $[a,b]$. If the property is satisfied in the interval, then there is no need to wait for time $b$ to affirm that the property is indeed verified.

This paper introduces preliminary works on the specification and verification of STL operators, using synchronous observers. The rest of the document focuses on discrete times, and non nested temporal operators. For example, STL properties like $\Box_{[a,b]} (\Diamond_{[c,d]} \varphi)$ with $\varphi$ an atomic proposition and $a$, $b$, $c$ and $d$ distinct times such that $a<b$ and $c<d$,  are excluded due to  the nested $\Diamond$ operator. 

Our main contribution concerns the formal verification of the correctness of STL operators. To this end, we provide a three-valued online STL semantics as well as the implementation of each STL operator in the synchronous language Lustre. The soundness of the implementation is expressed as a set of lemmas and automatically proved with the Kind2 model-checker. 

Section~\ref{sec:STL_intro} covers the preliminary concepts, including the Signal Temporal Logic, the synchronous language Lustre, the model checker Kind2, and an introduction to three-valued logic. We formalise an online semantics for STL operators in Section~\ref{sec:3VL_semantics}, and detail its Lustre implementation in Section~\ref{sec:Lustre}. Finally, Section~\ref{sec:Verif} describes the formal correction of the operators implementation.

\section{Preliminaries}\label{sec:STL_intro}

\subsection{Signal Temporal Logic}

Let $\mathbb{T}$ denote a set of discrete times such that $\mathbb{T} = \mathbb{N}$ and let $\mathcal{X}$ be a finite sets of signals. Let $a,b \in \mathbb{T}$ with $a < b$.
Without loss of generality, we assume that all signals are defined as functions in $\mathbb{T} \rightarrow \mathbb{R}$ from time to real values. To simplify notations, we denote the set of time $[t+a, t+b]$ as $t+[a,b]$. 

\begin{definition}[STL formal grammar]
    Let $\mu$ be an atomic predicate whose value is determined by the sign of a function of an underlying signal $x\in\mathcal{X}$, i.e., $\mu(t) \equiv \mu(x(t)) > 0$. Let $\varphi$, $\psi$ be STL formulas. STL formula $\varphi$ is defined inductively as:
        \vspace{-1em}
        
    \begin{displaymath}
    \varphi ::= \mu ~~|~~\neg \varphi ~~|~~ \varphi \land \psi ~~|~~ \varphi \lor \psi ~~|~~ \Box_{[a, b]} \psi ~~|~~ \Diamond_{[a, b]} \psi ~~|~~ \varphi ~\mathcal{U}_{[a, b]}~ \psi
    \end{displaymath}
\end{definition}

\begin{figure}
    \begin{eqnarray}
         (\mathcal{X}, t) \mymodels \mu &\Leftrightarrow&  \mu(t) \label{eq_notation_mu} \\
         (\mathcal{X}, t) \mymodels \neg\varphi &\Leftrightarrow& \neg ((\mathcal{X}, t) \mymodels \varphi) \label{eq_negation}\\
         (\mathcal{X}, t) \mymodels \varphi_1 \wedge \varphi_2 &\Leftrightarrow&  (\mathcal{X}, t) \mymodels \varphi_1
        \wedge (\mathcal{X}, t) \mymodels \varphi_2 \label{eq_and}\\
        (\mathcal{X}, t) \mymodels \varphi_1 \vee \varphi_2 &\Leftrightarrow&  (\mathcal{X}, t) \mymodels \varphi_1
        \vee (\mathcal{X}, t) \mymodels \varphi_2 \label{eq_or}\\
        (\mathcal{X}, t) \mymodels \varphi_1 ~\mathcal{U}_{[a,b]}~ \varphi_2 &\Leftrightarrow&  \exists t' \in t+[a, b]:  (\mathcal{X}, t') \mymodels \varphi_2
        \wedge \forall t'' \in [t, t']:(\mathcal{X}, t'') \mymodels \varphi_1 \label{eq_until}\\
        (\mathcal{X}, t) \mymodels \Diamond_{[a,b]}\varphi  &\:\Leftrightarrow&  \exists t' \in t+[a, b]: (\mathcal{X}, t') \mymodels \varphi \label{eq_diamond}\\
        (\mathcal{X}, t) \mymodels \Box_{[a,b]}\varphi &\:\Leftrightarrow&  \forall t' \in t+[a, b]: (\mathcal{X}, t') \mymodels \varphi \label{eq_box}
    \end{eqnarray}
    \caption{STL offline semantics}
    \label{fig:semantic_STL_offline}
\end{figure}

\begin{definition}[STL semantics]
The semantics of a formula $\varphi$ is defined at a time $t \in \mathbb{T}$ and for a set of signals $\mathcal{X}$ as $(\mathcal{X}, t) \mymodels \varphi$ as described in the \Cref{fig:semantic_STL_offline}. 

$\mu$ is evaluated locally, at time $t$ over the current values of the signals, Eq.~\eqref{eq_notation_mu}. Equation~\eqref{eq_negation} (Negation) is the logical negation of $\varphi$. Equation~\eqref{eq_and} (And) is the logical conjunction between $\varphi_1$ and $\varphi_2$. Equation~\eqref{eq_or} (Or) is the logical disjunction between $\varphi_1$ and $\varphi_2$.

It is worth mentioning that, in STL, all temporal operators have to be associated to a bounded, non-singleton time interval. Equation~\eqref{eq_until} (Until) describes a temporal operator that is satisfied if $\varphi_1$ holds from time $t$ until $\varphi_2$ becomes True within the time horizon $t + [a,b]$. Equation~\eqref{eq_diamond} (Eventually) describes a temporal operator that is satisfied if $\varphi$ is verified at least once within the time horizon $t + [a,b]$. Finally, Equation~\eqref{eq_box} (Always or Globally) describes a temporal operator that is satisfied  if $\varphi$ is always verified within the time horizon $t + [a,b]$. Note that the usual definitions of $\Diamond_{[a,b]}$ and $\Box_{[a,b]}$ based on $\mathcal{U}_{[a,b]}$ still apply:\vspace{-1em}
\begin{eqnarray}
    \Diamond_{[a,b]} \varphi &=& True ~\mathcal{U}_{[a,b]} ~ \varphi, \text{ and}\\
    \Box_{[a,b]}\varphi &=& \neg(\Diamond_{[a,b]} \neg\varphi).
\end{eqnarray}
\end{definition}

\begin{remark}
     While evaluation of predicates is performed at time $t$ in $(\mathcal{X}, t) \mymodels p \:\Leftrightarrow\:  \mu(t)$, all occurrences of time intervals $[a,b]$ in the definitions of $\mathcal{U}_{[a,b]}$, $\Box_{[a,b]}$ or $\Diamond_{[a,b]}$ are used to delay the current time $t$: $t+[a, b] = [t+a, t+b]$. These times $a$ and $b$ are then relative times while $t$ acts more as an absolute time.
\end{remark}

\subsection{Lustre}

 \begin{wraptable}{r}{.6\textwidth}
  \centering
\vspace{-1em}
  \begin{tabular}{r c l}
$td$ &::=&$ \mathbf{type}\ bt\ |\ \mathbf{type}\ t\ = enum\ \{\ C_i, ... \}$\\
$bt$ &::=& $real \ | \ bool \ |\ int \ |\ enum\_ident$\\
$d$ &::=& $\mathbf{node}\ f\ (p)\ \mathbf{returns}\ (p);$\\ && $\mathbf{vars}\ p\ \mathbf{let}\ D\ \mathbf{tel}$\\
$p$ &::=& $x : bt; ...; x : bt$\\
$D$ &::=& $pat = e; D\ |\ pat = e;$\\
$pat$ &::=& $x\ |\ (pat, ..., pat)$\\
$e$ &::=& $v\ |\ x\ |\ (e, ..., e)\ |\ e \rightarrow e\ |\ op (e, ..., e)$\\
&&$|\ \mathbf{if}\ e\ \mathbf{then}\ e\ \mathbf{else}\ e\ |\ \mathbf{pre}\ e$\\
$v$ &::=& $C\ |\ i$\\
\end{tabular}
  \caption{A subset of Lustre syntax}
  \label{fig:lustre_syntax}
  \vspace{-2em}
\end{wraptable} 
\textbf{Lustre}\cite{DBLP:conf/popl/CaspiPHP87} is a synchronous language
for modeling systems of synchronous reactive components.
 A Lustre program $L$ is a finite collection of nodes
$[N_0, N_1, \dots, N_m]$. The nodes satisfy the grammar described in
Table~\ref{fig:lustre_syntax} in which $td$ denotes type constructors,
including enumerated types, and $v$ either constants of enumerated
types $C$ or primitive constants such as integers $i$. 
Each node is declared
by the grammar construct $d$ of Table~\ref{fig:lustre_syntax}. 
A Lustre node $N$ transforms infinite streams of \textit{input} flows to streams of \textit{output} flows, with possible local variables denoting \textit{internal} flows. A notion of a symbolic ``abstract'' universal clock is used to model system progress. At each time step $k$, a node reads the value of each input stream and instantaneously computes and returns the value of each output stream. Note that all the equations of a node are computed at each time step. Therefore an if-then-else statement is purely functional and both of its branches are evaluated while only one of the computed value is returned.

\textbf{Stateful constructs.} Two important Lustre operators are the unary right-shift \texttt{pre} (for \texttt{pre}vious) operator and the binary initialization $\to$ (for followed-by) operator. 
Their semantics is as follows. 
For the operator \textit{Pre: }at first step $k=0$, \texttt{pre} \textit{p} is undefined, while for each step $k > 0$ it returns the value of \textit{p} at $k-1$. 
For the operator \textit{$\to$: }At step $k=0$, \textit{p} $\to$ \textit{q} returns the value of \textit{p} at $k=0$, while for $k>0$ it returns the value of \textit{q} at $k$ step.
  
For example, the Lustre equation $y = x_0 \to pre(u);$ will be defined for each time step $k$ by:
\begin{equation*}
    y(k) = 
    \left\{\begin{matrix} 
    x_0(0) &  \ \text{if } k = 0 \\ 
    u(k-1) &\  \text{if } k > 0
     \end{matrix}\right.
\end{equation*}

\subsection{Specifying and verifying assume-guarantee contracts with Kind2} \label{subsec_kind2}

The annotation language CoCoSpec~\cite{Champion2016} was proposed for Lustre models
 to lift
the notion of Hoare triple~\cite{DBLP:journals/cacm/Hoare69} and
Assume/Guarantee statements as dataflow contracts.
A contract is associated to a node and has only access to the input/output streams of that node. The body of a contract may contain a set of \texttt{assume} ($A$) and \texttt{guarantee} ($G$)
statements and \texttt{mode declarations}. Modes are named and consist of \texttt{require} ($R$) and \texttt{ensure} ($E$) statements. Assumes, guarantees, requires, and ensures are all Boolean expressions over streams. In particular, assumptions and requires are expressions over input streams, while guarantees and ensures are expressions over input/output streams. 
A synchronous observer corresponds to such a contract with only a guarantee statement. A node \textit{satisfies} a contract $C= (A, G')$ if it satisfies $Historically(A) \Rightarrow G'$, where $G'= G \cup\ \{R_i \Rightarrow E_i \}$ and $Historically(A)$  when A is true at all time.

Contracts can also define local flows, acting as \textit{ghost variables}. These potentially stateful flows can then be used in \texttt{guarantees} and \texttt{ensure} statements.

The following is an example of function \verb|timeab| in Lustre using a local contract \Cref{fig:Lustre_contract}. \verb|timeab| is a Lustre node indicating whether the current time is inside a given time interval $[a,b]$. It takes as inputs the integers $a$ and $b$, and returns a boolean value \verb|time| that is \verb|True| if the current time is inside $[a,b]$.
First line of the contract (line 3) defines a local variable \verb|clk| as an integer, which initially takes the value $0$ and is then incrementing at each time.
The \textit{assume} at line 4 indicates to model checker that it has to prove the Lustre node only in the cases where the condition $a \geq 0$ is satisfied. If another Lustre node is using \verb|timeab|, Kind-2 also checks that this node could not provide an input $a$ which runs counter to this assumption. Finally, Kind-2 must guarantee the equality, line 5, for all the inputs respecting the previous assumption, whatever the current time is. This equality compares the \verb|time| output to value of the specification clock in a valid interval $[a, b]$.
\verb|timeab| shall calculate the same output with an internal clock bounded at time $b$. So here, we verify that bounding the clock has no effect on the provided output.

\begin{figure}
    \begin{lustre}   
            node timeab (const a,b: int) returns (time: bool);
            (*@contract
            	  var clk : int = 0 -> 1 + pre clk;
                assume a >=0;
                guarantee time = (clk >= a and clk <= b);
            *)
    \end{lustre}
 \vspace{-1em}
    \caption{Example of a Lustre contract implementation}
    \label{fig:Lustre_contract}
\end{figure}

The Kind-2 model-checker~\cite{kind2} implements various SMT-based model-checking algorithms such as k-induction~\cite{kinduction} or IC3/PDR~\cite{ic3} and allows to verify contracts with respect to nodes.

\subsection{Three-valued logic}

\label{sec:3VL}

We present here the interest of three-valued logic, and introduce Kleene's three-valued logic, which we use in the next section to formalise an online version of STL operators.

For most tools, when performing monitoring of STL predicates, for a given value of simulation data, the trajectory is typically finite. It is produced by a simulation engine and stored in a data file. It is then loaded by the monitoring tool and analyzed with respect to the STL specification. In this offline setting, the final outcome indicates whether or not the input signal satisfies the specification. It is a boolean output.

Temporal operators are used to evaluate properties that change over time. Most of the time in these situations, we need to wait to decide whether a temporal property is satisfied or violated. Based on this observation, how to evaluate a property before being able to conclude, i.e. before the beginning of the time interval of a STL operator? Should we suppose that the operator is \verb|True| or \verb|False| before being able to decide? 

Let us consider a property $\Box_{[0,10]} P$, a set of signals $\mathcal{X}$ and an initial time $t_0$, e.g., $t_0=0$. We are interested in checking $(\mathcal{X}, t_0) \models \Box_{[0,10]} P$. Let us assume that we are given with a trace for $\mathcal{X}$ of length $l < 10$, e.g., $l = 8$, where the predicate $P$ is valid along the whole trace. What is the validity of such a predicate? On the one hand, it is always valid, but on the other hand, it has no real definition within the time $[l, 10]$. Indeed, $P$ could be false at time $t=9$ or $t=10$ and the property would be violated.

Since STL semantics requires all temporal connectors to be associated with bounded intervals, any STL predicate has a bounded horizon limit, after which it is always possible to determine the validity of a formula. We can then use this limit to evaluate temporal operators from it, considering that the values returned before may be irrelevant. But, in some case, the validity of the predicate can already be determined. In the previous example, if P is not valid at time $t=2$, we already know at this time that the operator will not be satisfied at the end of the time interval. Existing works regarding online semantics for STL~\cite{deshmukh2015robust} try to optimize the runtime evaluation of the predicate monitoring, detecting when one can conclude, positively or negatively. 

Rather than optimizing execution time based on binary logic, Łukasiewicz proposed a three-valued logic \cite{Lukasiewicz1970-LUKJLS}. This logic introduces a third truth-value, \verb|Unknown| ($\mathbf{U}$), describing values for which we are not yet able to conclude if the property is satisfied or violated. Later, Kleene proposes a strong logic of indeterminacy \cite{kleene52} similar to Łukasiewicz logic. The main difference lies in the returned value for implication. Kleene's approach states that $\mathbf{U} \Rightarrow \mathbf{U}$ is \verb|Unknown| while Łukasiewicz considers that $\mathbf{U} \Rightarrow \mathbf{U}$ should be \verb|True|. In our use of three-valued logic, we base ourselves on Kleene strong logic.

Table~\ref{tab:3vOpt} presents the truth tables showing the logical operations AND, OR, the logical implication, as well as the negation for Kleene's strong logic.

\begin{table}[t]
\small
    \begin{subtable}[ht]{0.22\textwidth}
        \centering
       \begin{tabular}{c|c|c|c|c}
   \toprule
    \multicolumn{2}{l|}{\multirow{2}{*}{A and B}} & \multicolumn{3}{c}{B}                               \\ \cline{3-5} 
    \multicolumn{2}{l|}{}                         & \multicolumn{1}{l|}{$\mathbf{F}$} & \multicolumn{1}{l|}{$\mathbf{U}$} & $\mathbf{T}$ \\ \midrule   \multicolumn{1}{l|}{\multirow{3}{*}{A}}  & $\textbf{F}$  & \multicolumn{1}{l|}{$\mathbf{F}$} & \multicolumn{1}{l|}{$\mathbf{F}$} & $\mathbf{F}$ \\ \cline{2-5} 
    \multicolumn{1}{l|}{}                    & $\mathbf{U}$  & \multicolumn{1}{l|}{$\mathbf{F}$} & \multicolumn{1}{l|}{$\mathbf{U}$} & $\mathbf{U}$ \\ \cline{2-5} 
    \multicolumn{1}{l|}{}                    & $\mathbf{T}$  & \multicolumn{1}{l|}{$\mathbf{F}$} & \multicolumn{1}{l|}{$\mathbf{U}$} & $\mathbf{T}$ \\ \bottomrule
    \end{tabular}
       \caption{AND Operation}
       \label{tab:and}
    \end{subtable}
    \hfill
    \begin{subtable}[ht]{0.22\textwidth}
        \centering
       \begin{tabular}{c|c|c|c|c}
    \toprule
    \multicolumn{2}{l|}{\multirow{2}{*}{A or B}} & \multicolumn{3}{c}{B}                               \\ \cline{3-5} 
    \multicolumn{2}{l|}{}                        & \multicolumn{1}{l|}{$\mathbf{F}$} & \multicolumn{1}{l|}{$\mathbf{U}$} & $\mathbf{T}$ \\ \midrule
    \multicolumn{1}{l|}{\multirow{3}{*}{A}}  & $\mathbf{F}$ & \multicolumn{1}{l|}{$\mathbf{F}$} & \multicolumn{1}{l|}{$\mathbf{U}$} & $\mathbf{T}$ \\ \cline{2-5} 
    \multicolumn{1}{l|}{}                    & $\mathbf{U}$ & \multicolumn{1}{l|}{$\mathbf{U}$} & \multicolumn{1}{l|}{$\mathbf{U}$} & $\mathbf{T}$ \\ \cline{2-5} 
    \multicolumn{1}{l|}{}                    & $\mathbf{T}$ & \multicolumn{1}{l|}{$\mathbf{T}$} & \multicolumn{1}{l|}{$\mathbf{T}$} & $\mathbf{T}$ \\ \bottomrule
    \end{tabular}
    
        \caption{OR Operation}
        \label{tab:or}
     \end{subtable}
     \hfill
         \begin{subtable}[ht]{0.22\textwidth}
           \centering
           \begin{tabular}{c|c}
           \toprule
             A & $\lnot A$\\
            \midrule
             $\mathbf{F}$ & $\mathbf{T}$ \\
             $\mathbf{U}$ & $\mathbf{U}$ \\
             $\mathbf{T}$ & $\mathbf{F}$ \\
            \bottomrule
           \end{tabular}
        \caption{Negation Operation}
        \label{tab:not}
     \end{subtable}
     \hfill
        \begin{subtable}[ht]{0.22\textwidth}
            \centering
           \begin{tabular}{c|c|c|c|c}
        \toprule
        \multicolumn{2}{l|}{\multirow{2}{*}{A $\Rightarrow$ B}} & \multicolumn{3}{c}{B}                               \\ \cline{3-5} 
        \multicolumn{2}{l|}{}                        & \multicolumn{1}{l|}{$\mathbf{F}$} & \multicolumn{1}{l|}{$\mathbf{U}$} & $\mathbf{T}$ \\ \midrule
        \multicolumn{1}{l|}{\multirow{3}{*}{A}}  & $\mathbf{F}$ & \multicolumn{1}{l|}{$\mathbf{T}$} & \multicolumn{1}{l|}{$\mathbf{T}$} & $\mathbf{T}$ \\ \cline{2-5} 
        \multicolumn{1}{l|}{}                    & $\mathbf{U}$ & \multicolumn{1}{l|}{$\mathbf{U}$} & \multicolumn{1}{l|}{$\mathbf{U}$} & $\mathbf{T}$ \\ \cline{2-5} 
        \multicolumn{1}{l|}{}                    & $\mathbf{T}$ & \multicolumn{1}{l|}{$\mathbf{F}$} & \multicolumn{1}{l|}{$\mathbf{U}$} & $\mathbf{T}$ \\ \bottomrule
        \end{tabular}
    
        \caption{Logical implication}
        \label{tab:implies}
     \end{subtable}
\caption{Truth Tables showing Kleene's 3-valued strong logic operations}
     \label{tab:3vOpt}
     \vspace{-1em}
\end{table}


\section{Online semantics for STL}
\label{sec:3VL_semantics}

To evaluate STL properties online, we rely on Kleene strong three-valued logic introduced in Table~\ref{tab:3vOpt}. In this section, we first introduce a way to obtain a three-valued output as proposed by Kleene, from two-valued outputs. Then, we provide an online and three-valued semantics for STL properties. 

\begin{definition}[Positive, negative and indeterminacy logics]
     Each STL temporal operator can be expressed in a three-valued form. To implement it, we define three new concepts:
    \begin{inparaenum}
        \item A \textit{Positive} logic $\mathbf{T}$ returning \verb|True| when the property is satisfied, and \verb|False| when it is yet undetermined or negative.
        \item A \textit{Negative} logic $\mathbf{F}$ that acts like an alarm to underline a negative result, which means that a statement returns \verb|True| when we are sure that the property is not satisfied, and it returns \verb|False| otherwise (undetermined property or satisfied property situations).
        \item An \textit{Indeterminacy} logic $\mathbf{U}$ highlighting situations where it is not yet possible to conclude about the satisfaction or violation of the property.
    \end{inparaenum}   
\end{definition}

\begin{definition}
    Let $\varphi$, $\varphi_1$ and $\varphi_2$ be STL properties. We denote by $\mathbf{T}^{~t}_\varphi$ (resp. $\mathbf{U}^{~t}_\varphi$ and $\mathbf{F}^{~t}_\varphi$) the evaluation of $(\mathcal{X},t) \mymodels \varphi$ according to the positive (resp. indeterminacy and negative) logic. We denote by $\mathbf{B}_\varphi$ the evaluation of $\varphi$ according to the offline implementation introduced in \Cref{fig:semantic_STL_offline}. Note that $\mathbf{B}_\varphi$ does not depend on time instant $t$.
\end{definition}

\begin{property}[Complete and pairwise distinct] \label{prop:complete_pairwise_dist}
    At any time instant $t$, exactly one of the three logic returns \verb|True| for a given property.
    \begin{align}
        \mathbf{T}^{~t}_\varphi \lor \mathbf{U}^{~t}_\varphi \lor \mathbf{F}^{~t}_\varphi & \quad \text{(completeness)} \label{eq:completeness} \\
        \lnot ((\mathbf{T}^{~t}_\varphi \land \mathbf{F}^{~t}_\varphi) \lor (\mathbf{T}^{~t}_\varphi \land \mathbf{U}^{~t}_\varphi) \lor (\mathbf{U}^{~t}_\varphi \land \mathbf{F}^{~t}_\varphi)) & \quad \text{(disjointness)} \label{eq:disjointness}
    \end{align}
\end{property}

\begin{remark}[Deduction of the output of the third logic]
    According to Property~\ref{prop:complete_pairwise_dist}, we only need the output of a given property in two of these three logics to determine its output in the last one. For example, if a property $\varphi$ returns \verb|False| in positive and negative logic, it means that $\varphi$ is still Unknown (\verb|True| in the indeterminacy logic). 
\end{remark}

\begin{remark}[Property determination]
    There exists an instant $t_d$ from which we cannot satisfy $\mathbf{U}^{t \geq t_d}_\varphi$. For a non-nested temporal operator evaluated on time interval $[a,b]$, $t_d$ corresponds at the latest to $t+b$. 
    \begin{equation}
        \exists t_d \leq t+b: \forall t' \geq t_d, \lnot \mathbf{U}^{~t'}_\varphi
    \end{equation}
\end{remark}

\begin{property} \label{prop:equiv_online_offline}
    From a specific time instant $t_f$, the offline and online results are similar. Thus, the outputs of the offline $\mathbf{B}_\varphi$ and online $\mathbf{T}^{~t_f}_\varphi$ versions are equivalent. In the same way, the negation of the offline operator is equivalent to the online negative version $\mathbf{F}^{~t_f}_\varphi$. For a non-nested temporal operator evaluated on time interval $[a,b]$, this time instant corresponds at the latest to $t+b$: 
    \begin{equation}
       \currenttime \geq t+b \implies ((\mathbf{B}_\varphi \iff \mathbf{T}^\currenttime_\varphi) \land (\lnot \mathbf{B}_\varphi \iff \mathbf{F}^\currenttime_\varphi)) \label{eq:equiv_online_offline}
    \end{equation}
\end{property}

\begin{property}[Immutability: Positive and negative logics are final]
    If a property is satisfied in the positive (resp. negative) logic, it will remain so in the future.
    \begin{align}
        \exists t \in \mathbb{T} : \mathbf{T}^{~t}_\varphi \implies \forall t' \geq t, \mathbf{T}^{~t'}_\varphi \label{eq:finality_T} \\
        \exists t \in \mathbb{T} : \mathbf{F}^{~t}_\varphi \implies \forall t' \geq t, \mathbf{F}^{~t'}_\varphi \label{eq:finality_F}
    \end{align}
\end{property}

From these three logics, we obtain easily a three-valued output. The property is:
\begin{inparaenum}
    \item \verb|True| in three-valued logic if it is \verb|True| in positive logic;
    \item \verb|False| in three-valued logic if it is \verb|True| in negative logic;
    \item \verb|Unknown| in three-valued logic if it is \verb|True| in indeterminacy logic.
\end{inparaenum}
    
Let us now characterize for each construct, the sufficient and necessary conditions to determine a positive, a negative, or a temporary indeterminate value.

In the case of a non-temporal property, the property validity can always be determined as either satisfied or violated. Let $\mu$ be an atomic proposition, ie. non temporal, and $t \in \mathbb{T}$: 
\begin{align}
    \forall \mathcal{X}, \forall t, \quad
    (\mathcal{X},t) \mymodels \mu \iff \mathbf{T}^{~t}_\mu \quad
    (\mathcal{X},t) \mymodels \lnot \mu \iff \mathbf{F}^{~t}_\mu  \quad
    \mathbf{U}_\mu = \bot   
\end{align}

If the property is a combination of multiple predicates based on logical operators ($\land, \lor, \implies, \neg\varphi, \ldots$), the validity is obtained using Kleene's three-valued strong logic presented in Table~\ref{tab:3vOpt}.

In the case of STL temporal operators as described in \Cref{fig:semantic_STL_offline}, we define a three-valued semantics describing when each operator is \verb|True|, \verb|False| or \verb|Unknown|. For positive and negative logics, we also provide an explicit version obtained by enumerating all the terms in the time horizon $t+a$ and $t+b$. The unknown explicit version for a property \verb|P| can be obtained by combining the positive and negative explicit versions :

\begin{flalign}
    \mathbf{U}^\currenttime_P \text{~(explicit)} \iff \big(\lnot \mathbf{T}^\currenttime_P \text{~(explicit)} \big) \land \big( \lnot \mathbf{F}^\currenttime_P  \text{~(explicit)} \big) 
\end{flalign}

Let us describe the positive, negative and indeterminate versions of each STL operator:

\begin{figure}
    \begin{flalign}
        \mathbf{T}^\currenttime_P \quad & \label{eq:diamond_3v_true} {\currenttime \geq t+a} \land \exists t' \in [t+a, min(\currenttime, t+b)]: (\mathcal{X}, t') \mymodels \varphi \\
        \mathbf{F}^\currenttime_P \quad &  \label{eq:diamond_3v_false} \currenttime \geq t+b \land \forall t' \in [t+a,t+b], (\mathcal{X}, t') \mymodels \lnot \varphi\\
        \mathbf{U}^\currenttime_P \quad & \label{eq:diamond_3v_unknown}
        (\currenttime < t+a) \lor (\currenttime < t+b \land \forall t' \in [t+a,\currenttime], (\mathcal{X}, t') \mymodels \lnot \varphi) \\
        \mathbf{T}^\currenttime_P \text{ (explicit)} \quad & \label{eq:diamond_3v_true_expl} \big((\mathcal{X},t+a) \models \varphi \big) \lor \big((\mathcal{X},t+a+1) \models \varphi \big)\lor ... \lor 
        \big( (\mathcal{X},t+b-1) \models \varphi \big) \lor \big( (\mathcal{X},t+b) \models \varphi \big) \\
        \mathbf{F}^\currenttime_P \text{ (explicit)} \quad & \label{eq:diamond_3v_false_expl} \big( (\mathcal{X},t+a) \models \lnot \varphi \big) \land \big( (\mathcal{X},t+a+1) \models \lnot \varphi \big) \land ... \land \notag \\
        & \big( (\mathcal{X},t+b-1) \models \lnot \varphi \big) \land \big( (\mathcal{X},t+b) \models \land \lnot \varphi \big)
    \end{flalign}\vspace{-2em}

\caption{Three-valued semantics of Eventually operator:  $P = \Diamond_{[a,b]} \varphi$}
\label{eventually_3v_semantics}
\end{figure}

\begin{figure}
    \begin{flalign}
        \mathbf{T}^\currenttime_P \quad &\label{eq:box_3v_true} \currenttime \geq t+b \land \forall t' \in [t+a,t+b], (\mathcal{X}, t') \mymodels \varphi  \\
        \mathbf{F}^\currenttime_P \quad &\label{eq:box_3v_false} \currenttime \geq t+a \land \exists t' \in [t+a, min(\currenttime, t+b)]: (\mathcal{X}, t') \mymodels \lnot \varphi\\
        \mathbf{U}^\currenttime_P \quad &\label{eq:box_3v_unknown}
        (\currenttime < t+a) \lor (\currenttime < t+b \land \forall t' \in [t+a,\currenttime], (\mathcal{X}, t') \mymodels \varphi) \\
        \mathbf{T}^\currenttime_P \text{ (explicit)} \quad & \label{eq:box_3v_true_expl} \big( (\mathcal{X},t+a) \models \varphi \big) \land \big( (\mathcal{X},t+a+1) \models \varphi \big)\land ... \land
        \big( (\mathcal{X},t+b -1) \models \varphi \big) \land \big( (\mathcal{X},t+b) \models \varphi \big) \\
        \mathbf{F}^\currenttime_P \text{ (explicit)} \quad & \label{eq:box_3v_false_expl} \big( (\mathcal{X},t+a) \models \lnot \varphi \big) \lor \big( (\mathcal{X},t+a+1) \models \lnot \varphi \big) \lor ... \lor \notag \\
        & \big( (\mathcal{X},t+b-1) \models \lnot \varphi \big) \lor \big( (\mathcal{X},t+b) \models \lnot \varphi \big)
    \end{flalign}\vspace{-2em}
\caption{Three-valued semantics of Always operator: $P = \Box_{[a,b]} \varphi$}
\label{always_3v_semantics}
\end{figure}

\paragraph{Eventually $\Diamond_{[a,b]} \varphi$ } (Fig.~\ref{eventually_3v_semantics}) In the positive logic, Eq.~\eqref{eq:diamond_3v_true}, we need to wait for time $t+a$, the beginning of the time interval, to have a chance to conclude positively if a valid condition has been observed. From time $t+b$, if the condition was not yet valid, the positive eventually operator always returns \verb|False|. For the negative logic, Eq.~\eqref{eq:diamond_3v_false}, invalidity requires to wait until the end of the time interval, otherwise one cannot conclude. Finally, the validity is unknown if we have not yet reached the end of the time interval but have not yet observed a suitable time, Eq.~\eqref{eq:diamond_3v_unknown}.

The explicit positive version \Cref{eq:diamond_3v_true_expl} is obtained by considering each instant between $t+a$ and $t+b$. One of these instants is supposed to satisfy the property. We use the disjunction between all the terms to check it. At the opposite, explicit negative version \Cref{eq:diamond_3v_false_expl} returns \verb|True| if all the terms between $t+a$ and $t+b$ satisfy $\lnot \varphi$. We therefore rely on the conjunction between all the terms. 

\paragraph{Always $\Box_{[a,b]} \varphi$ } (Fig.~\ref{always_3v_semantics}) In the positive logic, Eq.~\eqref{eq:box_3v_true}, similarly to the negative case of the eventually operator, one needs to wait until the end of the interval to claim validity. For the negative logic. Eq.~\eqref{eq:box_3v_false}, we detect invalidity as soon as we observe an invalid time, within the proper time interval. Unknown cases are either before the time interval or within it, if the property $\varphi$ is valid, up to now, Eq.~\eqref{eq:box_3v_unknown}.

The Always explicit positive version \Cref{eq:box_3v_true_expl} returns \verb|True| if each instant between $t+a$ and $t+b$ satisfies $\varphi$. Similarly to the explicit negative version of Eventually, we use the conjunction to verify this point. For the explicit negative version to return \verb|True|, it suffices that at one instant between $t+a$ and $t+b$, the property $\varphi$ is not satisfied. Thus, we check the disjunction of all the terms, searching if one of them violates $\varphi$.

\begin{figure}
    \begin{flalign}
       \mathbf{T}^\currenttime_P \quad &\label{eq:until_3v_true} \left( \currenttime \geq t+a \right) \land \left( \right. \exists t_1 \in [t+a, min(\currenttime, t+b)]: 
       (\mathcal{X}, t_1) \mymodels \varphi_2 \land \forall t_2 \in [t, t_1], (\mathcal{X}, t_2) \mymodels \varphi_1 \left. \right) \\[10pt]
       \mathbf{F}^\currenttime_P \quad &\label{eq:until_3v_false} 
        (\exists t_6 \in [t, min(\currenttime, t+a)]: (\mathcal{X}, t_6) \mymodels \lnot \varphi_1) \quad \lor \notag \\
        & (\currenttime \geq t+a \land \currenttime < t+b \land \exists t_7 \in [t+a, \currenttime] : (\mathcal{X}, t_7) \mymodels \lnot \varphi_1 \land \notag \\
        & \quad \lnot (\exists t_8 \in [t+a,\currenttime]:(\mathcal{X}, t_8) \mymodels \varphi_2 \land \forall t_9 \in [t, t_8], (\mathcal{X}, t_9) \mymodels \varphi_1)) \quad \lor \notag \\
        & (\currenttime \geq t+b \land \lnot (\exists t_{10} \in [t+a,t+b]:(\mathcal{X}, t_{10}) \mymodels \varphi_2 \land \forall t_{11} \in [t, t_{10}], (\mathcal{X}, t_{11}) \mymodels \varphi_1)) \\
        \mathbf{U}^\currenttime_P  \quad &\label{eq:until_3v_unknown} (\currenttime < t+a \land \forall t_3 \in [t, \currenttime], (\mathcal{X}, t_3) \mymodels \varphi_1) \quad \lor \notag \\
        & \big( \currenttime \geq t+a \land \currenttime < t+b \land \forall t_4 \in [t, \currenttime], (\mathcal{X}, t_4) \mymodels \varphi_1 \quad \land \forall t_5 \in [t+a, \currenttime], (\mathcal{X}, t_5) \mymodels \lnot \varphi_2 \big) \\
        \mathbf{T}^\currenttime_P \text{ (explicit)} \quad & \label{eq:until_3v_true_expl} \Big( \big( \bigwedge_{n=0}^{a} (\mathcal{X}, n) \mymodels \varphi_1 \big) \land \big( (\mathcal{X}, t+a) \mymodels \varphi_2 \big) \Big) \lor \Big( \big( \bigwedge_{n=0}^{a+1} (\mathcal{X}, n) \mymodels \varphi_1 \big) \land \big( (\mathcal{X}, t+a+1) \mymodels \varphi_2 \big) \Big) \lor ... \lor \notag \\
        & \Big( \big( \bigwedge_{n=0}^{b-1} (\mathcal{X}, n) \mymodels \varphi_1 \big) \land \big( (\mathcal{X}, t+b-1) \mymodels \varphi_2 \big) \Big) \lor \Big( \big( \bigwedge_{n=0}^{b} (\mathcal{X}, n) \mymodels \varphi_1 \big) \land \big( (\mathcal{X}, t+b) \mymodels \varphi_2 \big) \Big) \\
        \mathbf{F}^\currenttime_P \text{ (explicit)} \quad & \label{eq:until_3v_false_expl} \Big( \bigvee_{n=0}^{a} (\mathcal{X}, n) \mymodels \lnot \varphi_1 \Big) \lor \Big( \bigvee_{n_1=a+1}^{b} \big( (\mathcal{X}, n_1) \mymodels \lnot \varphi_1 \land (\bigwedge_{n_2=a}^{n_1} (\mathcal{X}, n_2-1) \mymodels \lnot \varphi_2) \big) \Big) \lor \notag \\
        & \Big( \bigwedge_{n=a}^{b} (\mathcal{X}, n) \mymodels \lnot \varphi_2 \Big)
    \end{flalign}\vspace{-2em}
\caption{Three-valued semantics of Until operator: $P = \varphi_1 \mathcal{U}_{[a,b]} \varphi_2$}
\label{until_3v_semantics}
\end{figure}
    
\paragraph{Until $\varphi_1 \mathcal{U}_{[a,b]} \varphi_2 $} (Fig.~\ref{until_3v_semantics}) Until operator is the most complex. We conclude positively when an event $\varphi_2$ occurred within the proper time interval, and until this moment $\varphi_1$ was always satisfied, Eq.~\eqref{eq:until_3v_true}. For the negative logic, there are multiple conditions that can lead to a violation of the property. First before the time interval, if $\varphi_1$ is not satisfied. Then inside the time interval, if $(\mathcal{X},\tau) \models \lnot \varphi_1$ before the moment when $(\mathcal{X},\tau) \models \varphi_2$, or if it was false before. Finally from $t+b$, if $\varphi_2$ is never reached inside the time interval or if it was false before, Eq.~\eqref{eq:until_3v_false}. About indeterminacy, we cannot yet conclude on the validity of the formula, if, for the moment the formula is neither validated nor violated. A first condition is that $\varphi_1$ holds from time $t$ until now. A second is that, at the current time $\currenttime$, we have not reach yet $t+a$ or we always have $\lnot \varphi_2$. These condition only apply before reaching the end of the time interval $t+b$, Eq.~\eqref{eq:until_3v_unknown}.

As the Until operator depends at the same time to the satisfaction of a property inside the time interval, and the satisfaction of another one before and inside the time interval, the explicit versions are less trivial to obtain than for others operators. For the explicit positive version to return \verb|True|, we need to satisfy the Until property at least once between $t+a$ and $t+b$, so we proceed by disjunction. Each term of the disjunction is satisfied only if $\varphi_1$ is satisfied from time $t$ until this time instant included (conjunction between all the terms between instant $t$ and this time instant) and $\varphi_2$ is satisfied at this moment. Note that time $t$ is represented by the $0$ value in the Until temporal referential, in the same way that instants $t+a$ or $t+b$ correspond to time $a$ or $b$ inside Until. As there are several ways of violated the Until operator, explicit false version of Until is built differently as others explicit versions. Indeed, we have a disjunction between the three possibilities to not satisfy the Until operator, as described above. We verify if $\varphi_1$ is not satisfied before or at time $t+a$ by relying on the disjunction between all the $\varphi_1$ terms from $t$ until $t+a$. Then, inside the time interval after time $t+a$, the property is violated with certainty if there exist a moment where $\varphi_1$ is not \verb|True|, and until the previous time, $\varphi_2$ was never satisfied. Indeed, if $\varphi_2$ was satisfied previously, either the property is satisfied, which means that explicit negative version must return \verb|False|; either the property was already violated before, so there exist another anterior time where $\varphi_1$ was not satisfied before $\varphi_2$ was satisfied. Finally, explicit negative version must return \verb|True| if $\varphi_2$ is never satisfy inside the time interval, which is studied by examining the conjunction of all the $\lnot \varphi_2$ terms between $t+a$ and $t+b$.


\section{Operators implementation strategy} \label{sec:Lustre}

Based on this online semantics, we propose an implementation of \textit{Eventually}, \textit{Always} and \textit{Until} in discrete time. We use the synchronous language Lustre. 
We recall that all the nodes are available at \therepository.

\paragraph{Useful constructs for the implementation}

First, we define the basic nodes needed to implement the temporal operators. 
Node \lustrein{min} returns the minimum value between two variables. 
Node \lustrein{exist(time:bool; prop: bool)} returns \verb|True| as soon as a property \verb|prop| has been satisfied during the time interval represented by \verb|time|. 
Node \lustrein{forall_a(time: bool ; prop: bool)} returns \verb|True| if a property \verb|prop| has always been \verb|True| during the time interval. All these nodes can easily be implemented in Lustre.

Regarding the implementation of the nodes detecting whether or not we are in the time slot $t+[a, b]$, and since we work with finite intervals, we can optimize our clock, preventing it from incrementing to infinity. 
We can limit the counter until value $b$, ensuring the absence of overflow.
We implement the node \lustrein{timeab}, that returns \verb|True| if the current time instant is inside the time interval, based on this bounded internal counter. As counter stops at $b$, end of the time interval is intercepted looking at the counter previous value. If it was already $b$, we know that we exceeded the end of the time interval. 

We are now able to implement our nodes for each version of each operator. In the case of the \textit{Positive} and \textit{Negative} versions, we want to stay as close as possible to the definition proposed in the Section~\ref{sec:3VL_semantics}. We take two liberties in order to optimise the memory management. First, for each operator, we define a bounded internal clock as described above. The same strategy as for the counter is used to determine the end of the time interval, comparing the previous value of the internal node counter with \verb|b|. Secondly, we want to have a bounded number of memories, not dependent on the trace-length or on the length of the time interval. We proceed as described in the literature \cite{donze_robust_2010}, by reusing the outputs obtained at the previous time instant to obtain the outputs at the current one. For example, here is the implementation of the Until False node: \Cref{fig:until_false_implementation}. Others \textit{Positive} and \textit{Negative} versions are obtained based on the same principle. 
\begin{figure}
    \centering
    \begin{lustre}
        node until_false (a,b: int ; phi1, phi2: bool) 
            returns (result_until_false: bool);
        var until_time: int;
        let
            -- internal clock 
            until_time = min(0 -> pre until_time + 1, b);
        
            -- init t=0 : until is violated if phi1 is false
            result_until_false = not phi1 ->
            
            -- violated if phi1 false before a
            ((until_time <= a) and (not phi1)) or
            
            -- violated if we are in the time interval,...
            (until_time > a and until_time <= b and 
              exist(timeab(a,b), not phi1) 
              -- and before this moment we never had
              and not (exist(timeab(a,b), 
              -- phi2 is true and until this moment phi1 is true.
              ((phi2) and forall_a(timeab(0,b),phi1))))) or
            
            -- violated if there is no instant in the time interval
            ((until_time >= b) and not (exist(timeab(a,b), 
            -- where phi2 is true and until this moment 
            -- phi1 is true
              ((phi2) and forall_a(timeab(0,b),phi1))))) or
            
            -- still violated if it was violated once in the past
            pre result_until_false;
        tel
    \end{lustre}
    \caption{Until Lustre node for the False version of the operator}
    \label{fig:until_false_implementation}
\end{figure}
Last, we deduce the \textit{Unknown} version from the \textit{Positive} and \textit{Negative} versions, as described in the Property~\ref{prop:complete_pairwise_dist}.

\begin{remark} \label{exclusivity_positive_negative_versions}
    The case where both \textit{Positive} and \textit{Negative} versions of the operator are \verb|True| at the same time is never supposed to happen and would result in an error. Indeed, it would mean that the property is both satisfied and violated, which is impossible. This result comes directly from Property~\ref{prop:complete_pairwise_dist}.
\end{remark}

Note that these implementations can only represent non-nested STL operators. That is to say that we only consider $\Diamond_{[a,b]}\varphi$, $\Box_{[a,b]}\varphi$ and $\varphi_1\mathcal{U}_{[a, b]}\varphi_2$ with $\varphi$, $\varphi_1$ and $\varphi_2$ being non-temporal predicates.

\section{Formal verification of STL operators} \label{sec:Verif}

In this section, we demonstrate by model checking that the operators implementation described in Section~\ref{sec:Lustre} corresponds to the given specification, as presented in Section~\ref{sec:3VL_semantics}. We first introduce the formalizing of each STL operator proof node for positive, negative and three-valued versions. Then, we present the use of Kind2 to concretely verify these proof nodes.

\subsection{Induction on time interval size}

To demonstrate the correctness of the positive and negative versions of the operators, we compare the outputs of our implementation proposition for each operator and an explicit equivalent, as provided in \Cref{eq:diamond_3v_true_expl,eq:diamond_3v_false_expl,eq:box_3v_true_expl,eq:box_3v_false_expl,eq:until_3v_true_expl,eq:until_3v_false_expl}. We remind that the explicit version is obtained by enumerating all the terms in the time horizon $t+[a,b]$. This allows to check directly the value of each term of the operator, and hence, to be sure to understand the obtained output. We have to show that our implementation and the explicit one are equivalent for any time interval. We prove this property by strong structural induction on the time interval size. 

We proceed as follows. In a first time, we demonstrate that a statement is true for the smallest possible STL time interval, cf. base case of Eq.~\eqref{eq:base_case}. Then, we demonstrate that if the statement is true for a given time interval size, it is also true when we increase the size interval by 1, cf. Eq.~\eqref{eq:inductive_case}. By verifying these two properties, we demonstrate the correctness of our operators for all intervals $[a,b]$ such that  $a, b \in \mathbb{T} \land a < b$.

\paragraph{Base case: $[a, a+1]$}
Let $\mathbf{Op}$ be a version of a temporal operator, and $\mathbf{Op\_exp}$ its explicit representation as described in \Cref{sec:3VL_semantics}. 
\begin{equation}
    \forall a \in \mathbb{T}, \mathbf{Op}_{[a,a+1]} \varphi \iff \mathbf{Op\_exp}_{[a, a+1]} \varphi \label{eq:base_case}
\end{equation}

For the base case proof, we create a new Lustre node \lustrein{P_at_k}, cf. Fig.~\ref{fig:P_at_k} that checks if a property is satisfied at a specific time or was satisfied before. This allows us to implement the explicit case. Let us take the example of the \textit{Positive} version of the Eventually. For the $[a, a+1]$ time interval, its explicit version is Eq.~\eqref{eq:expl_a_ap1} and its implementation corresponds to the lines 13 and 14 of the~\Cref{fig:ev_true_proof_node}
\begin{equation} \label{eq:expl_a_ap1}
    \Diamond_{[a,a+1]} \varphi \equiv 
    \left((\mathcal{X}, a) \models \varphi\right) \lor \left((\mathcal{X}, a+1) \models \varphi\right)
\end{equation}

\begin{figure}
    \begin{lustre}
        node P_at_k (const k: int; clk:int; P:bool) 
            returns (ok: bool);
        let
          ok = if clk = k then P else (false -> pre ok);
        tel
    \end{lustre}
    \caption{$P\_at\_k$ Lustre node}
    \label{fig:P_at_k}
\end{figure}

\paragraph{Inductive case: $[a, b+1]$}

Let $\mathbf{Op}$ be a version of a temporal operator and $\mathbf{Op\_exp}$ its explicit representation as described in Section~\ref{sec:3VL_semantics}. 
\begin{align}
    (\mathbf{Op}_{[a,b]} \varphi \iff \mathbf{Op\_exp}_{[a, b]} \varphi) \implies
    (\mathbf{Op}_{[a,b+1]} \varphi \iff \mathbf{Op\_exp}_{[a, b+1]} \varphi) \label{eq:inductive_case}
\end{align}

In our implementation, $\mathbf{Op\_exp}_{[a, b+1]}$ is obtained thanks to the previous value of $\mathbf{Op}_{[a,b]}$, that we assume equivalent to $\mathbf{Op\_exp}_{[a, b]}$. For example, inductive case of the explicit version of Eventually True operator is obtained as described in Eq.~\eqref{eq:inductive_case_impl} and its implementation corresponds to lines 16 and 17 of~\Cref{fig:ev_true_proof_node}
\begin{align} 
\Diamond_{[a,b+1]} \varphi \equiv 
\left(\Diamond_{[a,b]} \varphi\right) \lor \left((\mathcal{X}, b+1) \models \varphi\right) \label{eq:inductive_case_impl}
\end{align}

\medskip
To concretely check these basic and inductive cases, we use the Kind2 model checker, cf~\Cref{subsec_kind2}. For each positive and negative version of STL operators, we express the base and inductive case as two properties, ie. two lemmas inside a contract to guarantee basic and inductive case, cf. \Cref{fig:ev_true_proof_node}. 

\begin{figure}
    \centering
    \begin{lustre}
        returns (base_case,ind_case: bool);
        (*@contract
           assume a<b and a>=0;
           guarantee base_case;
           guarantee ind_case;
        *)
        var clk : int;
            output_ev_true, output_ev_true_bp1: bool;
        let
          clk = 0 -> 1 + pre clk;
          output_ev_true = eventually_true (a,b,phi);
          output_ev_true_bp1 = eventually_true (a,b+1,phi);
          base_case = (b=a+1) =>
            (output_ev_true = P_at_k(a,clk,phi) or P_at_k(a+1,clk,phi));
                  
          ind_case =
            (output_ev_true_bp1 = (output_ev_true or P_at_k(b+1,clk,phi)));
        tel
    \end{lustre}\vspace{-1em}
    \caption{Eventually True proof node}
    \label{fig:ev_true_proof_node}
\end{figure}

Finally, to obtain a three-valued output, we need to encode the result on two booleans. We combine the positive and negative outputs - previously verified - to determine the state of the operator. According to Property~\ref{prop:complete_pairwise_dist}, Unknown is obtained if Positive and Negative outputs return \verb|False| at the same time. As a complementary check, we ensure inside a contract that Positive and Negative versions are mutually exclusive as mentioned in the Remark~\ref{exclusivity_positive_negative_versions}. Figure~\ref{fig:ev_3v} summarizes the implementation of this final node in Lustre.

\begin{figure}
    \centering
    \begin{lustre}
        node always_3v(const a,b:int; phi: bool)
          returns (output_ev_true, output_ev_false: bool);
        (*@contract 
          assume a<b and a>=0;
          -- their are mutually exclusive
          guarantee not (output_ev_true and output_ev_false);
        *)
        let  
            output_ev_true = eventually_true(a,b,phi);
            output_ev_false = eventually_false(a,b,phi);
        tel
    \end{lustre}\vspace{-1em}
    \caption{Three-valued Eventually node in Lustre}
    \label{fig:ev_3v}
\end{figure}

\subsection{Using Kind2 as a theorem prover}

Each of the three temporal operators is defined in a separate file. They all rely on basic nodes mentioned in Sect.~\ref{sec:Lustre}, in which only \lustrein{timeab} is fitted with a contract. The following tables summarize all contract elements automatically proved by Kind2 model-checker. We recall that Kind2 relies on different model-checking algorithms that are executed in parallel. The method that succeeds first interrupt the proof process. In the table, \emph{PDR} stands for Property-Direct-Reachability \cite{ic3} while \emph{k-induction} specifies the number of steps of the k-induction process used to conclude.
In both methods, Kind2 produces subproblems that are solved using Z3 \cite{z3}.

As mentioned above, each operator \lustrein{op} is defined using two underlying nodes \lustrein{op_false} and \lustrein{op_true} as well as a node \lustrein{op_3v} that reconstruct the three-valued output. The nodes \lustrein{op_false} and \lustrein{op_true} are not directly   associated to a contract but their soundness is expressed through the validity of another node: respectively \lustrein{proof_op_false} and \lustrein{proof_op_true}. These nodes are defining the base 
and inductive 
cases and associated to the main contract (cf. Fig.~\ref{fig:ev_true_proof_node}). Last, the final node \lustrein{op_3v} is only fitted with an extra contract guarantying disjunctiveness of the output (cf. Fig.~\ref{fig:ev_3v} encoding Eq.~\eqref{eq:disjointness}).   

\begin{table}
\small
\centering
\begin{tabular}{|l|c|c|c|}
\hline
Node name & \# Property & Method & Proof time\\\hline\hline
\footnotesize\verb!timeab!& assume & PDR & 0.339s\\
& assume & induction & 0.351s\\
& guarantee & PDR & 2.618s\\
\hline
\footnotesize\verb!eventually_true! & - & - & -\\ \hline 
\footnotesize\verb!proof_ev_true!& assume & PDR & 0.653s\\
& assume & 2-induction & 0.713s\\
& guarantee & 2-induction & 26.778s\\
& guarantee & 2-induction & 29.631s\\
\hline
\footnotesize\verb!eventually_false! & - & - & -\\ \hline 
\footnotesize\verb!proof_ev_false!& guarantee & PDR & 37.215s\\
& guarantee & PDR & 37.215s\\
\hline
\footnotesize\verb!eventually_3v!& assume & PDR & 0.677s\\
& assume & 2-induction & 0.688s\\
& guarantee & 2-induction & 0.688s\\
& guarantee & 2-induction & 13.216s\\
& guarantee & 2-induction & 18.855s\\
\hline
\hline
\end{tabular}
    \caption{Experiments for operator Eventually.}
\end{table}

\begin{table}
\small
\centering
\begin{tabular}{|l|c|c|c|}
\hline
Node name & \# Property & Method & Proof time\\\hline\hline
\footnotesize\verb!timeab!& assume & PDR & 0.432s\\
& assume & 2-induction & 0.454s\\
& guarantee & PDR & 2.855s\\
\hline 
\footnotesize\verb!until_true! & - & - & -\\ \hline 
\footnotesize\verb!proof_until_true!& assume & PDR & 0.537s\\
& assume & 2-induction & 0.595s\\
& guarantee & 2-induction & 6.347s\\
& guarantee & 2-induction & 93.526s\\
& guarantee & 2-induction & 161.614s\\
\hline
\footnotesize\verb!until_false! & - & - & -\\ \hline 
\footnotesize\verb!proof_until_false!& assume & induction & 0.898s\\
& assume & induction & 0.898s\\
& assume & induction & 0.898s\\
& guarantee & 2-induction & 606.067s\\
& guarantee & PDR & 1605.403s\\
\hline
\footnotesize\verb!until_3v!& guarantee & PDR & 34.530s\\
\hline
\hline
\end{tabular}
   \caption{Experiments for operator Until.}
\end{table}

\begin{table}
\small
\centering
\begin{tabular}{|l|c|c|c|}
\hline
Node name & \# Property & Method & Proof time\\\hline\hline
\footnotesize\verb!timeab!& assume & PDR & 0.400s\\
& assume & induction & 0.425s\\
& guarantee & PDR & 3.375s\\
\hline
\footnotesize\verb!always_true! & - & - & -\\ \hline 
\footnotesize\verb!proof_alw_true!& guarantee & 2-induction & 76.633s\\
& guarantee & 2-induction & 96.686s\\
\hline
\footnotesize\verb!always_false! & - & - & -\\ \hline 
\footnotesize\verb!proof_alw_false!& guarantee & 2-induction & 19.407s\\
& guarantee & PDR & 27.540s\\
\hline
\footnotesize\verb!always_3v!& guarantee & PDR & 12.854s\\
\hline
\hline
\end{tabular}
    \caption{Experiments for operator Always.}
\end{table}

Experiments were run with kind2 v2.0.0-7-gdcc7f6f on a 1,2 GHz Quad-Core Intel Core i7 with 16 GB of RAM. To build the table, each node is analyzed independently, but a quicker analysis of each file can be performed with all nodes analyzed at once. Note also results with the same execution time such as the elements of the node \lustrein{eventually_3v}. Typically, in this case, they denote properties that were proved together k-inductive by the algorithm. We observe something similar with PDR for node \lustrein{proof_ev_false}.

As a last remark, we have to say that, because of the parallel architecture of Kind2, it is difficult to obtain perfect reproductibility of the results. For example, one can observe that the runtime of the validity proof of the simple node \lustrein{timeab_tmp} varies slightly between experiments while it is the exact same node. The difference can also appear in the number of unrolling of the k-induction engine.

\section{Discussions and conclusion}

\paragraph{Related Works.} The use of three-valued logic has already been explored in the context of temporal logic, particularly in LTL, with the same division used in this paper: one value indicating the certainty of satisfaction of a property, another indicating the certainty of violation of a property, and a final value representing indeterminacy~\cite{bauer_monitoring_2006,ho_online_2014}. 

Formal verification of STL properties has also been studied. Roehm et al.~\cite{roehm_stl_2016} propose to check STL properties on reach sequences, using hybrid model checking algorithms such as Cora~\cite{althoff_introduction_2015} or SpaceEx~\cite{FrehseLGDCRLRGDM11}. A first step consists in the transformation of STL properties into their \textit{reachset temporal logic} (RTL) equivalent. This transformation comes close to the explicit development of each operator that we described in \Cref{sec:3VL_semantics}, requiring potentially a large set of memories. 

Moreover, several examples of algorithms and online implementation of STL properties have been produced, using a finite number of memories, cf~\cite{maler_monitoring_2013,donze_robust_2010}. Thus, \cite{donze_robust_2010} proposes an algorithm for quantitative online STL implementation, and show on different examples the time-saving benefits of using their online method compared to the offline one. Balsini et al.~\cite{balsini_generation_2017} propose a qualitative online implementation of STL in Simulink, which nevertheless has some limitations. In particular, since three-valued logic is not used in this implementation, we cannot be sure whether a property has been satisfied or violated until the end of execution. These proposals go further than ours, allowing operators to be nested, sometime with some limitations like \cite{balsini_generation_2017} that can only contain one operator inside another. However the soundness of the encoding is not formally proven.

\paragraph{Conclusion.} In this paper, we propose an online discrete implementation of the STL semantics, in the continuity of Balsini's work~\cite{balsini_generation_2017}. Our contribution is twofold. First, we proposed an implementation based on Kleene's three-valued logic in order to be able to represent indeterminacy. Second, we formally demonstrated the soundness of our implementation, proving the validity of each operator with respect to its semantics.

Our approach was the following: we first defined the online STL semantics, and used it to build each STL operator as a synchronous observer in the Lustre language. Finally, we formally demonstrated the correctness of their implementation, using the Kind2 model-checker. We proceed by induction on the size of temporal intervals. We succeed to demonstrate all the proof objectives for each temporal operator implemented. 

\paragraph{Future Work.} 
They are mainly two directions to continue this work. A first one is to apply these operators to models and see how model-checkers such as Kind2 can verify properties or produce counter-examples. For example revisiting the use case of Roehm et al.~\cite{roehm_stl_2016}. The other direction is to extend the set of STL formulas that can be encoded in our framework. While Balsini et al.~\cite{balsini_generation_2017} proposed a similar encoding (but without proof) of nested operators with restricted form and up to two levels, we would like to lift the restrictions and deal with more general formulas. The notion of propagation delays introduced in Kempa et al.~\cite{DBLP:conf/formats/KempaZJZR20} could also lead to an efficient encoding with memories, also associated with proof of the implementation.

\section{Acknowledgment}

The authors would like to thank the  Institute for Cybersecurity in Occitania (ICO) for partially funding this work.

\nocite{*}
\bibliographystyle{eptcs}
\bibliography{biblio}
\end{document}